\documentclass[11pt,twoside]{article}

\usepackage{graphicx}
\usepackage{asp2006}
\usepackage{epsf}
\usepackage{psfig}
\usepackage{lscape}

\begin{document}
\title{Lyman Break Galaxies in the VLT/FORS2 spectroscopic campaign in the
  GOODS-S field} 
\author{E. Vanzella$^{1}$, S. Cristiani$^1$, M. Dickinson$^2$, M. 
Giavalisco$^3$,  K. Lee$^3$, M Nonino $^1$, P. Rosati $^4$,  and the GOODS team}
\affil{$^1$INAF - Trieste Astronomical Observatory, via G.B. Tiepolo 11,
  40131, Trieste, ITALY} 
\affil{$^2$National Optical Astronomy Obs., P.O. Box 26732, Tucson, AZ 85726}
\affil{$^3$Space Telescope Science Institute, 3700 San Martin Drive,
  Baltimore, MD 21218} 
\affil{$^4$ESO Garching, Karl-Scwarzschild-Strasse 2, D-85748 Garching bei
  M{\"u}nchen, Germany}

\begin{abstract}
We present initial results from our ongoing campaign of spectroscopic
identifications of Lyman--break galaxies (LBGs) at $z\sim 4$, 5 and 6 with
FORS2 at the ESO VLT. 

\end{abstract}

\section{Introduction}

The Lyman--break selection technique (Guhathakurta et al. 1991; Steidel et
al. 1996; Giavalisco 2002; Giavalisco et al. 2004) remains one of the most
efficient way to build large and well controlled samples of star--forming
galaxies at $z>2$, suitable for a variety of statistical studies, such as
luminosity function, clustering and the source of cosmic reionization. Recent
panchromatic deep surveys such as the Great Observatories Origins Deep Survey
(GOODS, Giavalisco et al. 2004) have made possible to extend the
identification of LBGs at redshifts as high as $z\sim 6.5$ (Dickinson et
al. 2004; Bunker et al. 2004; Bouwens et al. 2006). The relative simplicity of
modeling the inherent selection effects of the Lyman--break technique as a
function of the galaxy properties (e.g. the dispersion of their UV SED) has
also made possible to measure the evolution of key properties such as
morphology, luminosity function, star--formation rate, stellar mass and AGN
activity of LBGs over a broad redshift range (for an overview of the GOODS
project see Renzini et al. 2002; Giavalisco \& Dickinson 2003; Giavalisco et
al. 2004a, and {\it http://www.stsci.edu/science/goods/}).

Here we report on preliminary results of an ongoing campaign of spectroscopic
identifications of Lyman--break galaxies up to $z\sim 6.5$ carried out in the
GOODS-S field with the FORS2 spectrograph at the ESO-VLT. We have secured so
far 103 identifications the redshift range 3.5-6.3 (Vanzella et al. 2005;
2006a; 2006b in prep.), currently the largest spectroscopic sample of LBGs in
these redshifts. In the following the ACS F435W,
F606W, F775W, and F850LP filters are denoted as $B$, $V$ , $i$ and $z^\prime$,
and magnitudes are in AB system.

\begin{figure}[!ht]
\includegraphics[width=14cm,height=5cm]{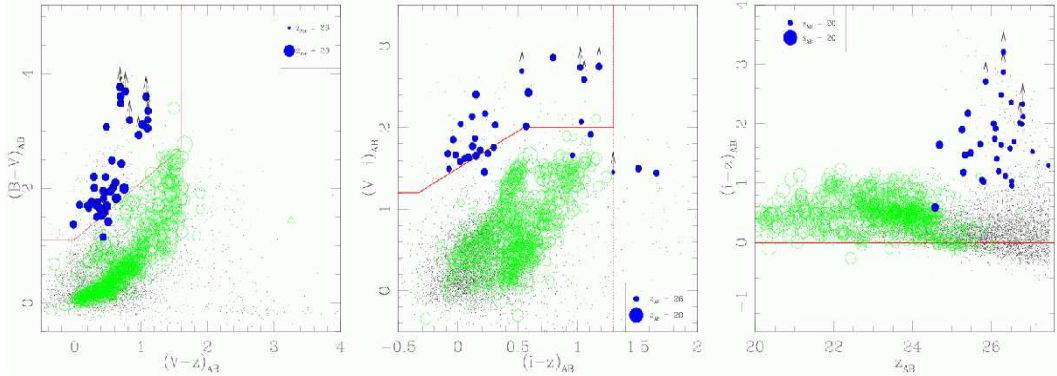}
\caption{Selection diagrams of $B$, $V$ and $i$-drop galaxies (from left to
 right).  The lines outline the regions of the selection. The black dots are
 colors of sources down to $z^\prime$=27.5. Open (green) and filled (blue)
 circles are sources with $z$$<$3.0 and 3.0$<$$z$$<$4.4 ($B$-drop), $z$$<$4.4
 and 4.4$<$$z$$<$5.4 ($V$-drop) and $z$$<$5.4 and $z$$>$5.4 ($i$-drop). The size
 of the symbols scale with the $z^\prime$ magnitude.}
\label{fig1}
\end{figure}
%
\begin{figure}[!ht]
\begin{center}
\rotatebox{0}{
\includegraphics[width=13cm,height=7cm]{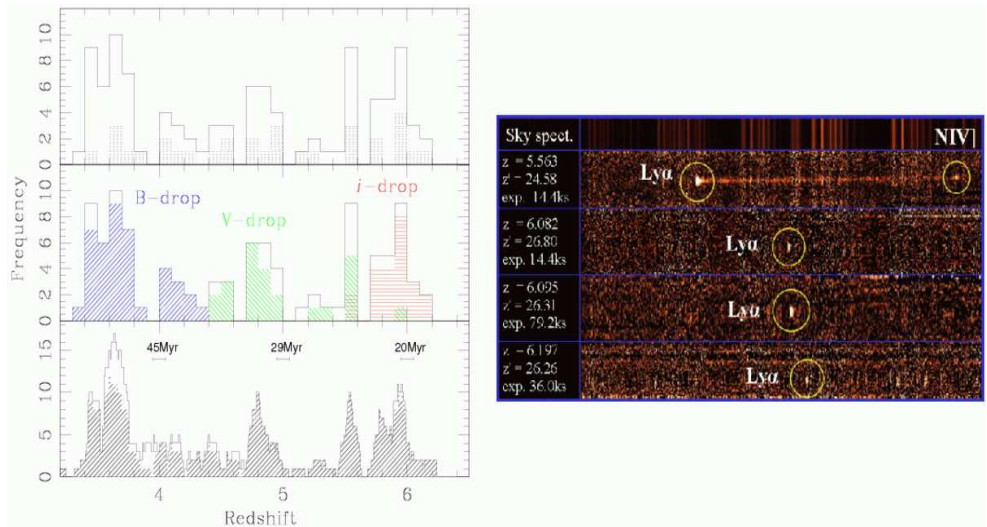}
}
\end{center}
\caption{\emph{Left:} Redshift distribution of the current GOODS-S LBGs spectroscopic 
samples. In the upper panel the dotted area represents the sources with
lower spectral quality (QF=``C''). Middle panel shows the redshift
distribution (continuum line) with the highlighted categories $B$, $V$ and
$i$-drops. Bottom panel shows the redshift distribution calculated in a
finer redshift bin, the shaded region is the FORS2 spectroscopic
sample, while the continuum line histogram includes additional redshift 
from the literature 
({\it  http://www.eso.org/science/goods/spectroscopy/CDFS$\_$Mastercat/}).
\emph{Right:} Examples of the FORS2 2-D spectra of $i$--band dropouts with 
Ly$\alpha$ in emission.}
\label{fig2} 
\end{figure} 

\section{Photometric selection of LBGs in the GOODS-S field}\label{nil:sec2}

We selected the targets of our spectroscopic observations from the samples of
$B$, $V$ and $i$--band ``dropouts'', at mean redshifts
$<z>\sim 3.8$, $4.8$ and $5.8$, respectively, obtained applying the color
equations by Giavalisco et al. (2004b) to the v1.0 release of the GOOODS ACS
catalogs. We have also included targets in close proximity of the boundary of
the selection window to test the effect of photometric scatter and measure the
efficiency of the selection criteria and the contamination by lower--redshift
interlopers.  Figure~\ref{fig1} shows the color-color diagrams and the LBGs 
with spectroscopic identifications. 

%
\begin{figure}[!ht]
\begin{center}
\includegraphics[width=13cm,height=5cm]{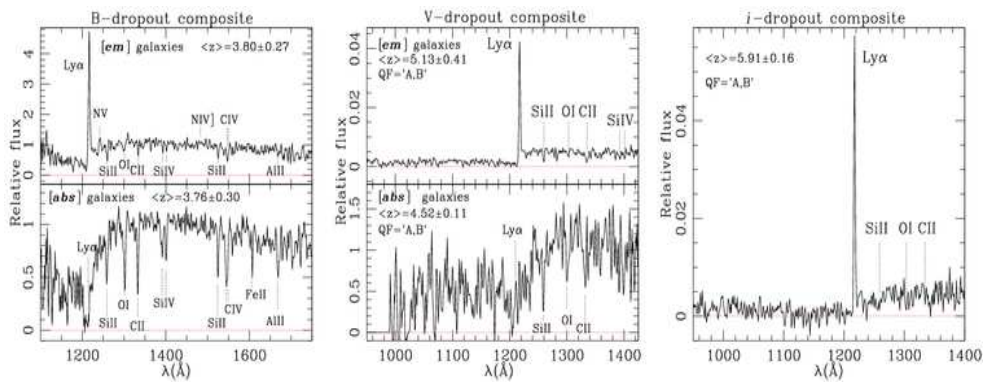}
\end{center}
\caption{From left to right, composite spectra of $B$, $V$ and $i$ drop
galaxies. The Ly$\alpha$ line, the typical stellar and interstellar absorption
lines and the break of the continuum blueward the Ly$\alpha$ line are
evident.}\label{fig3}
\end{figure}
%

We found the success rate of the redshift identifications to be $\approx 70
\%$ for all three categories of LBGs, and the efficiency of the selection for
source with redshift is 97.6$\%$, 86.2$\%$ and 79.3$\%$ for the $B$, $V$ and $i$
drops, respectively. In total 11 interlopers have been found, 10 of them have
been confirmed to be stars and one is a galaxy at redshift 1.3 with [OII] in
emission and evident Balmer break (originally selected as a $V$-drop). The 
current spectroscopic completeness of our $B$, $V$ and $i$--band dropout samples
is $\sim 5\%$, 14$\%$, 30$\%$ down to $z^\prime$=26.5.

Figure~\ref{fig2} (left) shows the redshift distribution of the current samples
LBGs. There is some overlap around redshift $\sim$4.5 between the $B$ and
$V$--band dropout samples and at $z\sim$5.5 between the $V$ and $i$--band
ones. The observed average and standard deviation of the $B$, $V$ and $i$--band
sample redshift distributions, as well as their shape, are in general good
agreement with the predicted ones computed according to the method outlined by
Giavalisco et al. (2004b).  The Figure~\ref{fig2} (left, lower panel) shows the
redshift distribution calculated in finer redshift bins, $dz=0.1$, with the  
overdensity at $z=5.9$ noted by Malhotra et al. (2005) clearly visible. 
\begin{figure}[!ht]
\centering
\rotatebox{0}{
\includegraphics[width=7cm,height=5cm]{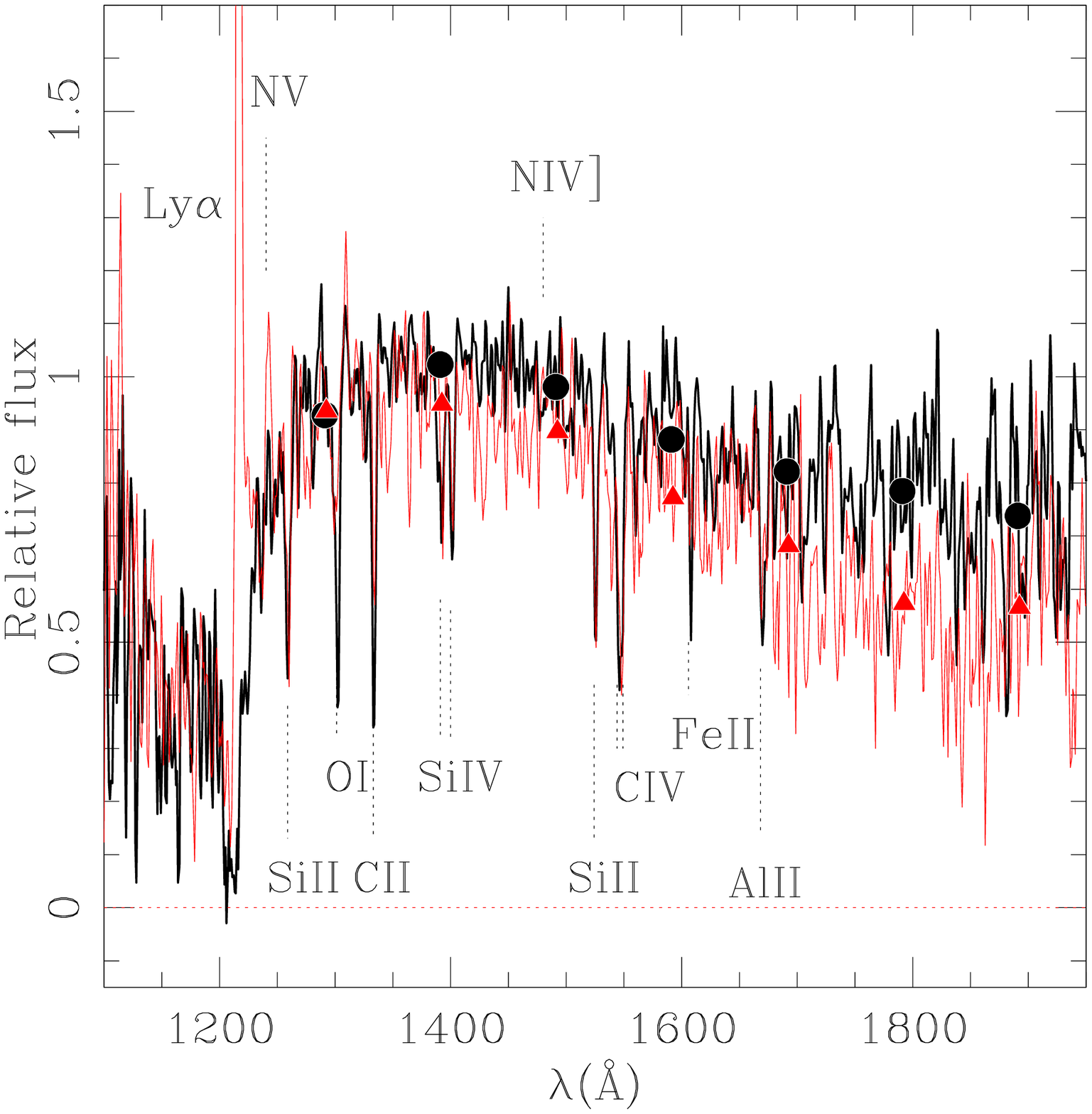}
\includegraphics[width=6cm,height=5cm]{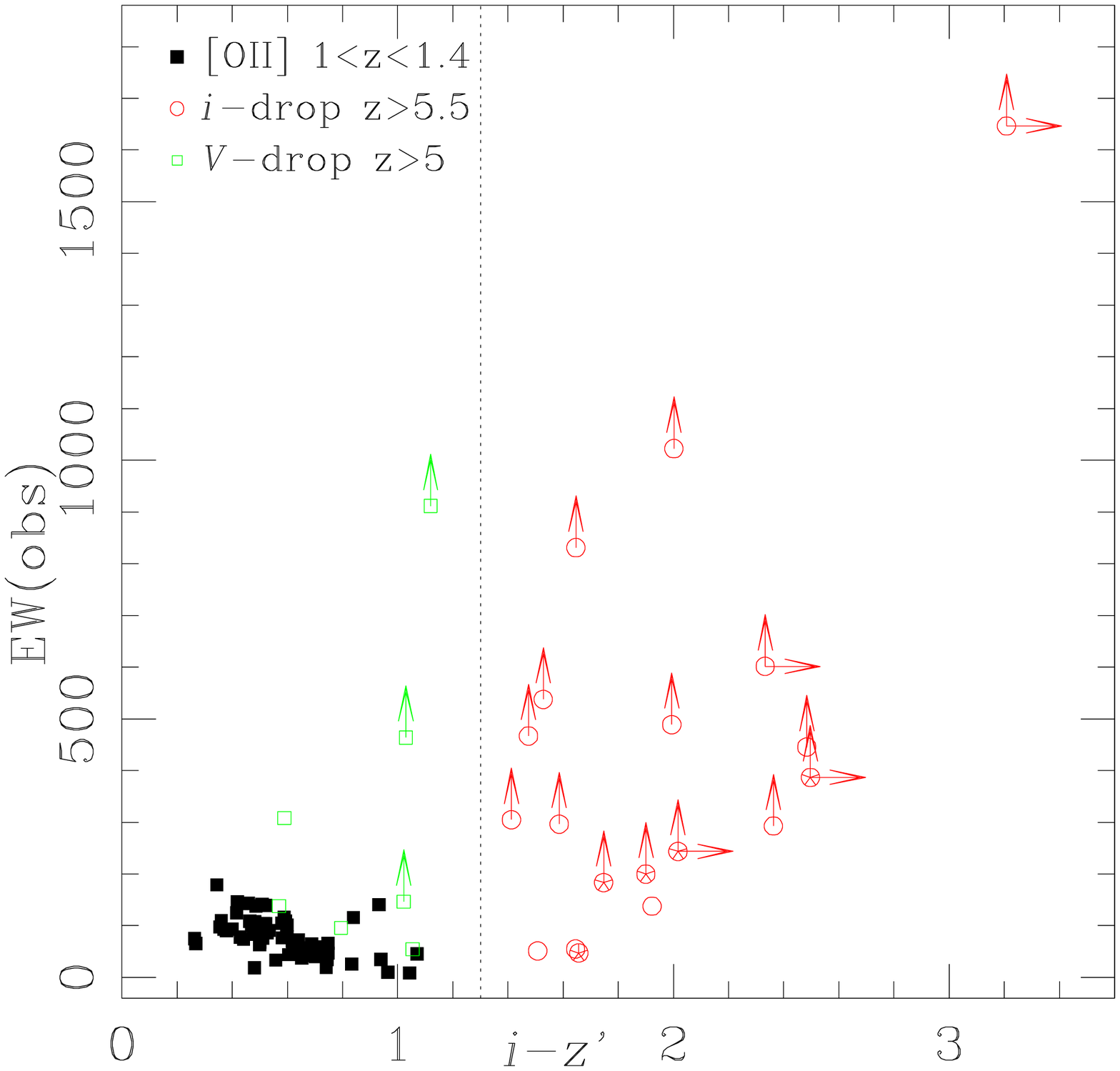}
}
\caption{\emph{Left:} 
Comparison between composite spectra normalized at 1300\AA~of the $B$-drop
galaxies with and without the Ly$\alpha$ emission line. Circles and triangles 
are the median values of the continuum computed in bins of 100\AA. 
\emph{Right:} Observed EW([OII]) of a sample of galaxies at 1$<$z$<$1.4
compared to the observed EW of the $i$--band dropout sample.}
\label{fig4}
\end{figure}

\section{Spectroscopic properties of LBGs in the GOODS-S field}\label{ev:sec2}

The quality of the spectra in our three samples varies, depending on the
luminosity and redshift of the sources; however, trends are easily
recognized, which show a close similarity of properties with LBGs observed at
$z\sim 3$ (Shapley et al. 2003). The typical rest--frame UV features observed
are the HI Ly$\alpha$ line, seen either in emission, absorption, or a
combination of both; low-ionization resonance interstellar metal lines such as
SiII $\lambda$1260, OI$\lambda$1302 + SiII $\lambda$1304, C II $\lambda$ 1334,
SiII $\lambda$1526. FeII $\lambda$1608, and AlII $\lambda$1670, which are
associated with the neutral interstellar medium, and high-ionization metal
lines such as Si IV $\lambda\lambda$1393,1402 and C IV
$\lambda\lambda$1548,1550 associated with ionized interstellar gas an P-Cygni 
stellar wind features.

Figure~\ref{fig3} shows composite spectra of the three samples of LBGs. For
$B$ and $V$--band dropouts spectra with the Ly$\alpha$ line in emission
(``emitters'') and absorption (``absorbers'') were stacked separately. The
sample of spectroscopically identified $i$--band dropouts, and hence the
composite spectrum, only includes ``emitters'', since the current sensitivity
precludes spectroscopic identifications from absorption features in such faint
candidates. Figure~\ref{fig4} (left) compares the stacked spectra of $B$--band dropout
absorbers and emitters, showing that the former have a systematically redder
UV continuum and stronger interstellar absorption lines than the latter, a
fact also observed in lower--redshift samples at $2<z<3$ (Shapley et
al. 2003). 

Most (but not all) $i$--band dropouts identifications are based on one
emission line only (see right panel of Figure~\ref{fig2} as an example), 
which we intepret as Ly$\alpha$, because of the line's
asymmetric profile, which we observe every time the S/N ratio is
sufficient. This is clearly visible in the stacked spectrum in Figure~\ref{fig3},
together with the sharp discontinuity of the continuum at the wavelength of
the feature, showing that the stacked spectrum is indeed dominated by galaxies
at $z\sim 6$. To further investigate the possibility that low--redshift
interlopers, such as galaxies with [OII] emission at $z\sim$ 1.1--1.4, might
contaminate the sample, we have compared the observed equivalent width and
($i$-$z$) color of the $i$--band dropout spectroscopic samples with that of
[OII] galaxies independently confirmed at $1.1<z<1.4$. 
Figure~\ref{fig4} (right) shows that the real [OII] galaxies are too blue and
the equivalent width of the [OII] emission line is too small compared to the
same quantities of the $i$--band dropouts, for a significant contamination to
be present.

\begin{figure}[!ht]
\centering
\rotatebox{0}{
 \includegraphics[width=6cm,height=6cm]{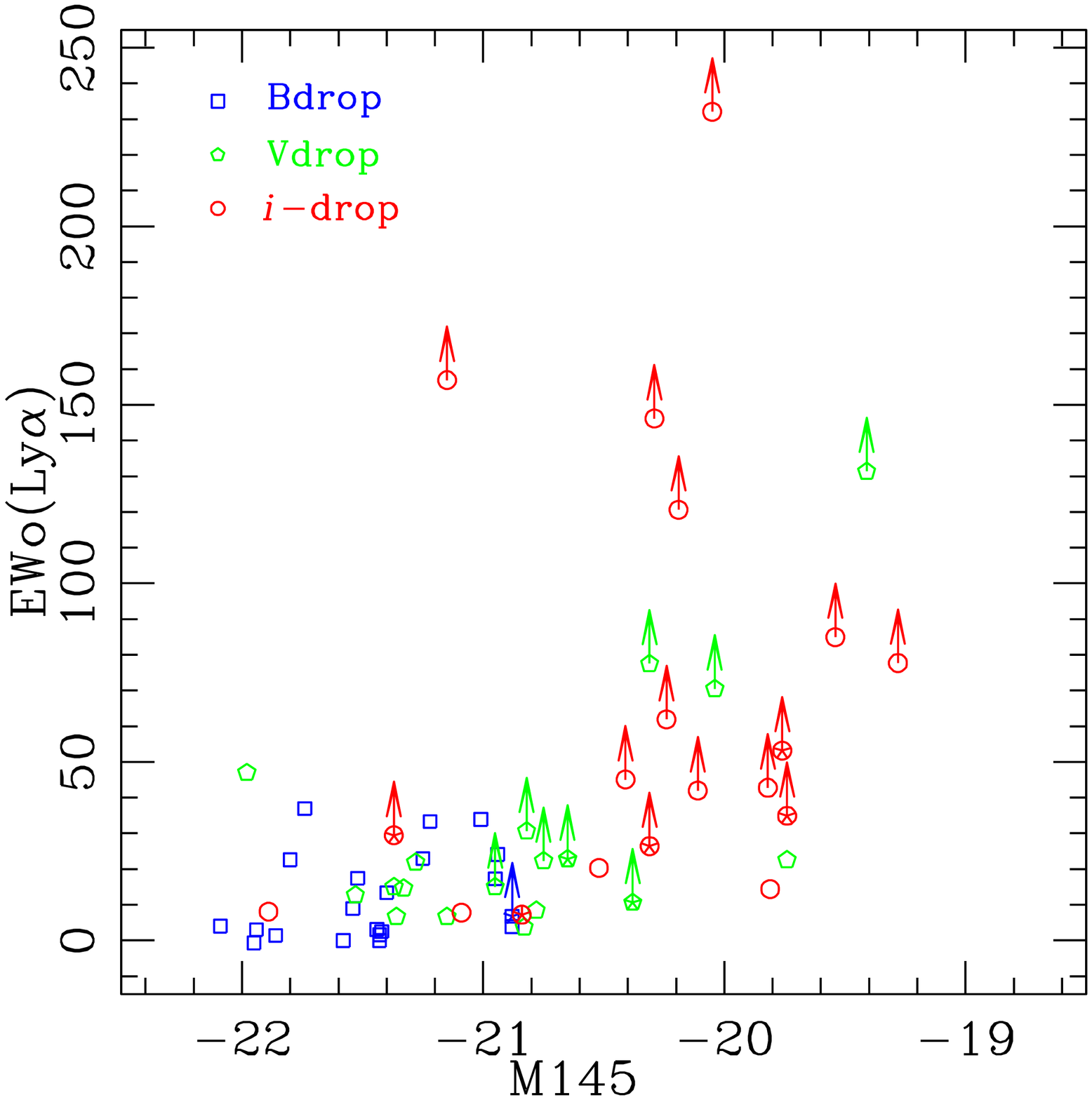}
 \includegraphics[width=6cm,height=6cm]{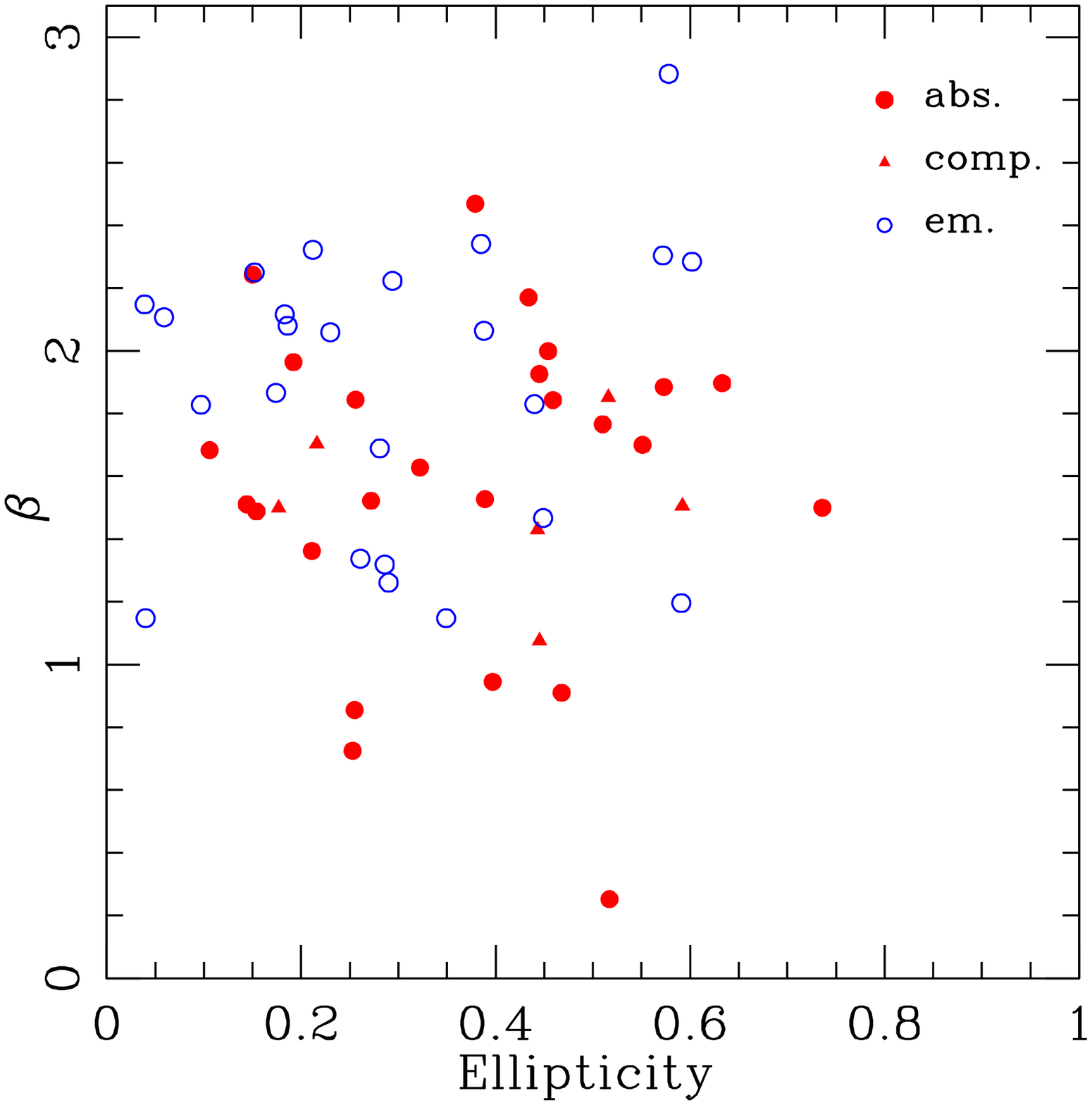}
 }
\caption{\emph{Left:} Rest frame Ly$\alpha$ EW of LBGs as a function of the UV
luminosity.  \emph{Right:} UV spectral index $\beta$ mesured in the
rest--frame 1700--2500 \AA\ and the ellipticity of the rest--frame UV
morphology}\label{fig5}
\end{figure}
The right panel of Figure~\ref{fig5} plots the UV spectral index $\beta$
($F_{\lambda}=\lambda^{-\beta}$) measured in the rest--frame 1500--1800 \AA\
from the $i$-$z^{\prime}$ color versus the ellipticity of the rest--frame UV
morphology for both ``emitters'' (blue) and ``absorbers'' (red) of the $B$--band
dropout sample. No correlation is observed between these quantitites for
either spectral ``types'' (suggesting that optically thick disks are not
predominant in the sample), but we find that the median $\beta$ of the
emitters, $\beta_{em}=2.1$, is larger than that of the absorbers,
$\beta_{abs}=1.7$, in agreement with the comparison of the stacked spectra.
This is probably evidence that ``emitters'' are on average younger and/or less
dust--obscured than ``absorbers''.

We also observe a velocity offset between the Ly$\alpha$ and the interstellar
absorption lines, in the sense that the Ly$\alpha$ is more redshifted than the
ISM by a velocity difference ($v_{em}-v_{abs}$)$\ge$ 300 and 400~km/sec for
the $B$ and $V$--band dropouts, respectively. This is in quantitative
agreement with what reported at lower redshifts (e.g. Shapley et al. 2003),
suggesting a similarity in the energetics of large--scale gas outflows.

Figure~\ref{fig5} (left) shows the rest--frame equivalent width the Ly$\alpha$ emission
line of the three dropout samples as a function of the continuum luminosity
density at $\lambda=1450$ \AA. Value range from a few \AA\ up to 
$\sim 250$ \AA, although for faint sources at z$>$5 we only measure lower
limits. The fact that large equivalent widths are observed only among
the fainter and more distant galaxies is an obvious selection effect; however, 
there clearly is a deficiency of strong Ly$\alpha$ emission lines among the
brightest and closest ones. A similar trend has also been observed by Ando et
al. (2006) in a mixed sample of LBGs and Ly$\alpha$ emitters as redshift
$5<z<6$. This may be evidence of different formation history and/or of 
ISM properties in bright and faint galaxies.

\begin{figure}[!ht]
\centering
\rotatebox{0}{
 \includegraphics[width=6cm,height=6cm]{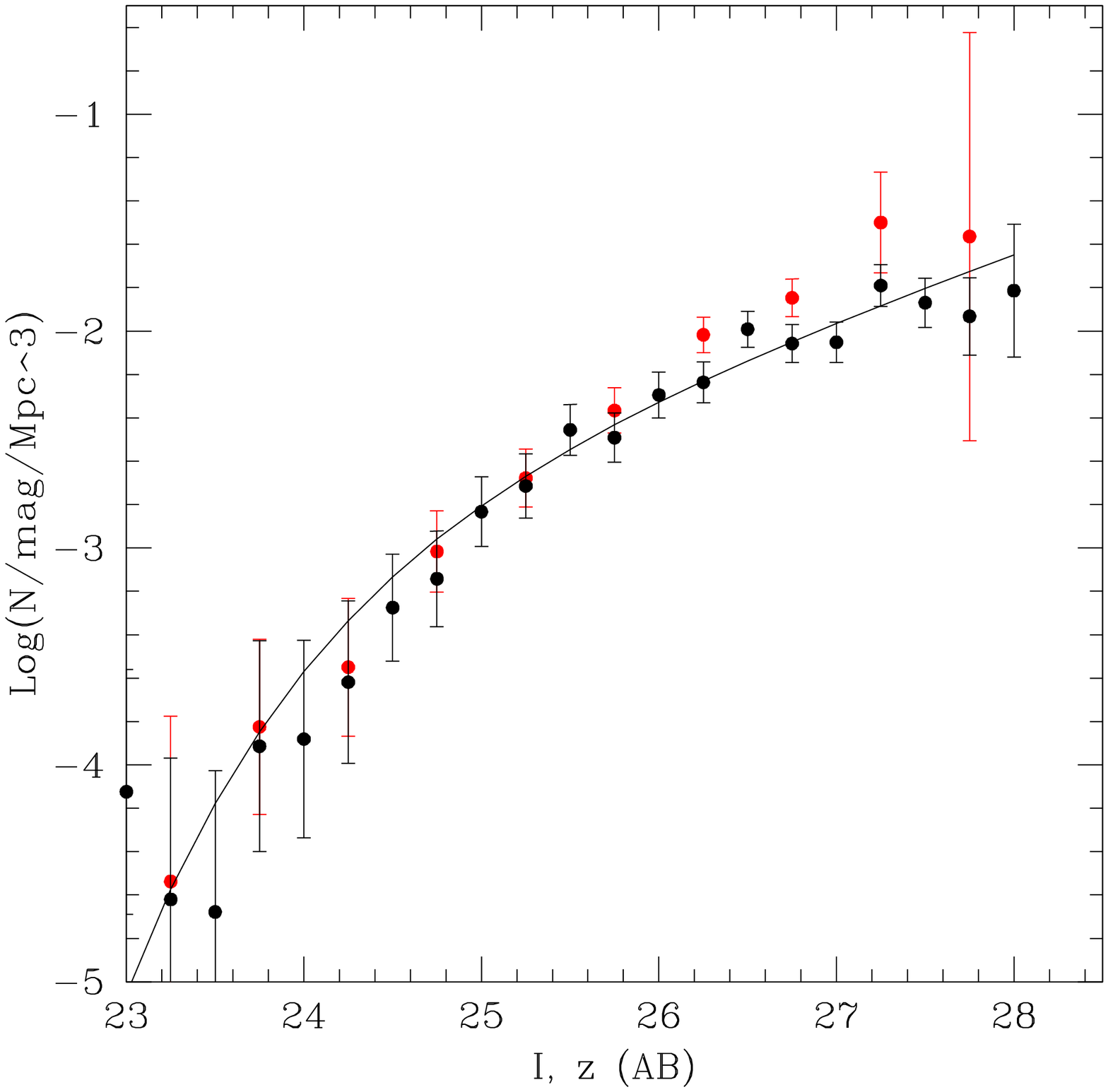} 
 \includegraphics[width=6cm,height=6cm]{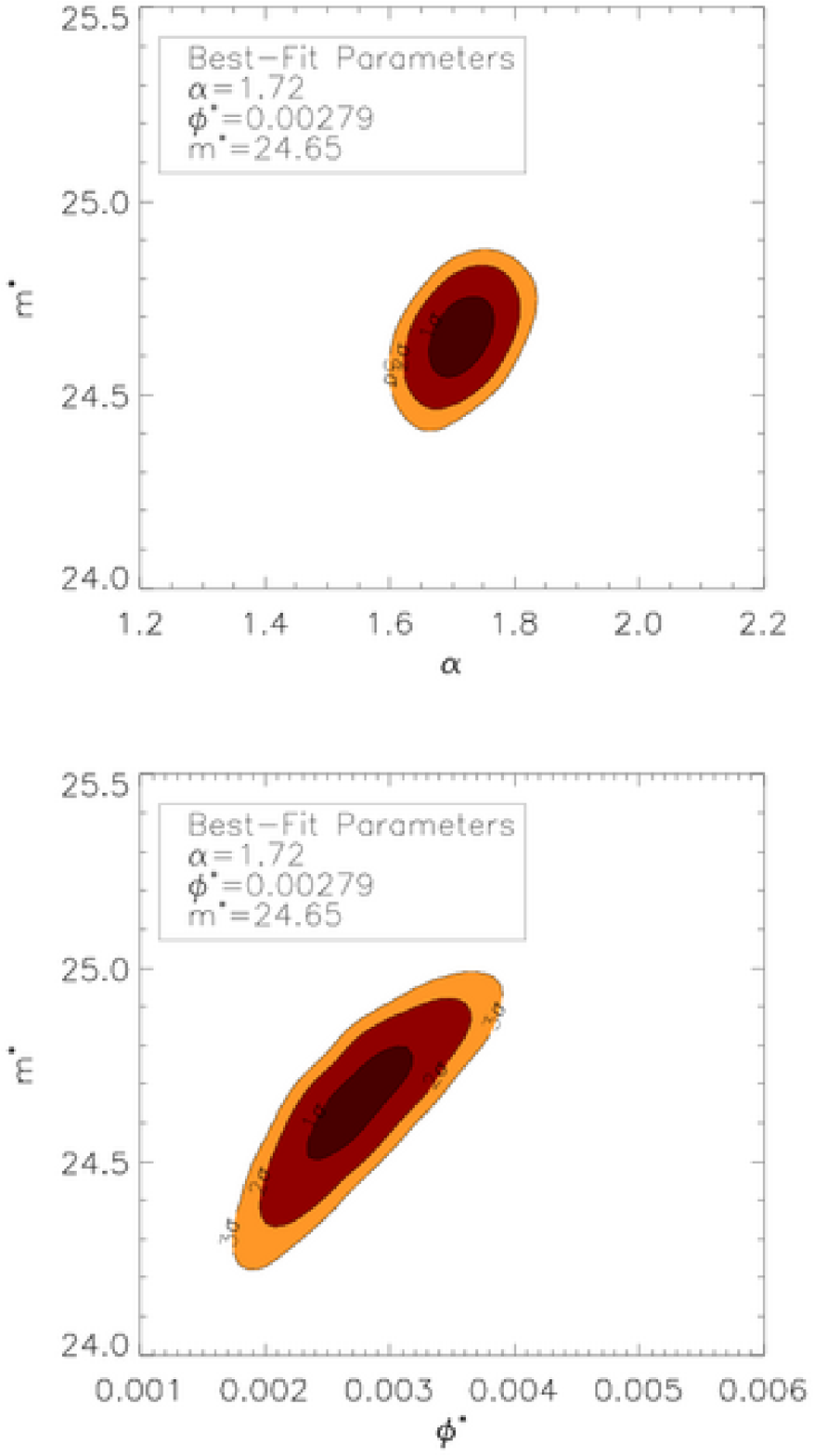} 
 }
\caption{LF of the GOODS $B$ and $V$--band
dropouts together with the best--fit Schechter function and the uncertainties
of its parameters.}\label{fig6}
\end{figure}

In Figure~\ref{fig6} we also show the luminosity function of $B$--band
dropouts from the two GOODS fields (black dots) and COSMOS (red dots) together
with the GOODS best--fit Schechter function and the uncertainties of its
parameters from a series of upcoming papers (Vanzella et al.; Giavalisco et
al.; Lee et al. in prep.). The corresponding rest--frame $\lambda=1500$
\AA\ absolute magnitude is $M^*_{B}=-21.3\pm 0.2$. A similar measure from the
$V$--band dropout samples yields $\phi^*_{V}=2.86\pm 0.01\times 10^{-3}$
Mpc$^{-3}$, $M^*_{V}=-21.3\pm 0.4$ and $\alpha_{V}=1.61\pm 0.15$. Finally,
from samples of $i$--band dropouts extracted from GOODS and the UDF, Bouwens 
et al. (2006) report $\phi^*_{i}=2.02\pm 0.25\times 10^{-3}$ Mpc$^{-3}$,
$M^*_{i}=-20.25\pm 0.5$ and $\alpha_{V}=1.73 \pm 0.5$, suggesting that the
evolution of the UV luminosity function of LBGs over the redshift range
$4<z<6$ is characterized by a relatively constant value of the faint end slope
$\alpha\sim-1.6$ and volume density $\phi^*$, together with 
a gradual dimming of the characteristic luminosity $M^*$ by $\sim 1$ mag.

\acknowledgements 
E.V. acknowledge financial contribution from contract ASI--INAF I/023/05/0.

\end{document}